\documentclass[aps,prl,reprint,groupedaddress]{revtex4-1}
\pdfoutput=1
\usepackage[T1]{fontenc}
\usepackage[utf8]{inputenc}
\usepackage{graphicx}
\usepackage{amsmath}
\usepackage{amssymb}
\usepackage{pifont}
\usepackage{float}

\date{\today} 

\begin{document}
\title{The Influence of Molecular Architecture on the Dynamics of H-Bonded Supramolecular Structures in Phenyl-Propanols}
\author{Till Böhmer}
\email{tboehmer@fkp.tu-darmstadt.de}
\author{Jan Gabriel}
\author{Timo Richter}
\author{Florian Pabst}
\author{Thomas Blochowicz}
\email{thomas.blochowicz@physik.tu-darmstadt.de}
\affiliation{Institut für Festkörperphysik, Technische Universität Darmstadt, 64289 Darmstadt, Germany.}

\begin{abstract}
  The relaxation behaviour of monohydroxy alcohols (monoalcohols) in broadband dielectric spectroscopy (BDS) is usually dominated by the Debye process. This process is regarded as a signature of the dynamics of transient supramolecular structures formed by H-bonding. In phenyl propanols the steric hindrance of the phenyl ring is assumed to influence chain formation and thereby to decrease or even suppress the intensity of the Debye process. In the present paper we study this effect in a systematic series of structural isomers of phenyl-1-propanol in comparison with 1-propanol. It turns out that by combining BDS, Photon Correlation Spectroscopy (PCS) and calorimetry the dynamics of supramolecular structures can be uncovered. While light scattering spectra show the same spectral shape of the main relaxation for all investigated monoalcohols, the dielectric spectra differ in the Debye contribution. Thus it becomes possible for the first time to unambiguously disentangle both relaxation modes in the dielectric spectra. It turns out that the Debye relaxation gets weaker the closer the position of phenyl ring is to the hydroxy group, in accordance with the analysis of the Kirkwood-Fröhlich correlation factor.  Even in 1-phenyl-1-propanol, which has the phenyl group attached at the closest position to the hydroxy group, we can separate a Debye-contribution in the dielectric spectrum. From this we conclude that hydrogen bonds are not generally suppressed by the increased steric hindrance of the phenyl ring, but rather an equilibrium of ring and chain-like structures is shifted towards ring-like shapes on shifting the phenyl ring closer to the hydroxy group. Moreover, the shape of the $\alpha$-relaxation as monitored by PCS and BDS remains unaffected by the degree of hydrogen bonding and is the same among the investigated alcohols.  
\end{abstract}

\maketitle 


\section{Introduction}

Molecular dynamics of hydrogen bonding systems is key for an understanding of a broad variety of biological and technologically relevant systems. For instance, protein folding and the DNA alpha helix formation are mediated through formation and breaking of hydrogen bonds (H-bonds).

To achieve a fundamental understanding of H-bonding materials, monohydroxy alcohols (monoalcohols) can serve as model systems due to their simplicity with one hydroxy group per molecule. As these substances are easily supercooled, molecular dynamics can be traced over a wide frequency range by dynamically sensitive techniques like broadband dielectric spectroscopy (BDS) and depolarized dynamic light scattering. 

Basically, BDS probes the reorientational motion of permanent molecular dipole moments\,\cite{Boettcher:1978b}.
Usually the slowest relaxation process observed by BDS in non-H-bonding supercooled liquids is the $\alpha$-process, which is related to the viscosity of the liquid and the structural relaxation. However, for many monoalcohols (e.g. 1-propanol) the picture is different: At relaxation times clearly slower than the $\alpha$-process, another strong relaxation appears in the dielectric spectra, which usually is not broadened like typical relaxation processes in supercooled liquids, but can be described by one single relaxation time and is therefore referred to as Debye process\,\cite{Boehmer:2014a,Hansen:1997}.

In monoalcohols this process is usually understood in the framework of a transient chain model\,\cite{Gainaru:2010a}: By H-bonding chain-like supramolecular structures are formed, which can either show reorientational  diffusive motion as a whole or  attachment or detachment of molecules to and from the chain. These structures possess a large overall dipole moment parallel to the end-to-end vector of the transient chain and move on a slower timescale than the $\alpha$-relaxation. 

The view is, among other things, supported by the static dielectric permittivity, which in monoalcohols usually deviates strongly from the values predicted by the Onsager equation. Such deviations result from orientational cross-correlations of neighboring dipole moments and can be quantified by the Kirkwood-Fröhlich correlation factor\,\cite{Kremer:2002a}:
\begin{equation}
g_\text{k}=1+\frac{1}{\mu^2}\left\langle \boldsymbol{\mu}_i(0)\cdot\sum_{j\neq i}\boldsymbol{\mu}_j(0) \right\rangle.
\end{equation}
Most monoalcohols display $g_\text{k}>1$ and therefore positive cross-correlations have to exist as in the case of chain-like supramolecular structures. 

Of course, the molecular architecture has a decisive impact on the formation of supramolecular structures, as was previously shown by several different studies, e.g. by dielectric and shear mechanical investigations of isomeric series of octanols\,\cite{Dannhauser:1968a,Dannhauser:1968b,Hecksher:2014}. In the mentioned studies monoalcohols with both $g_\text{k}>1$ and $g_\text{k}<1$ were identified. This was explained by Dannhauser et al. using a model that involves the occurrence of ring-like and chain-like H-bonded suprastructures, the ratio of which depends on temperature, pressure and molecular architecture. The appearance of predominantly ring-like structures would then lead to $g_\text{k}<1$ and even a temperature dependent transition from predominantly ring-like to chain-like structures, resulting in a transition from $g_\text{k}<1$ to $g_\text{k}>1$, was reported for 5-methyl-3-heptanol \cite{Dannhauser:1968b}. 

More recently it was found that closed loop-like structures can open up into linear chains under certain conditions like high electric fields \cite{Singh:2012,Young-Gonzales:2016} or high pressure \cite{Pawlus:2013}, where an increase of the Debye process was reported.

Nevertheless, it is still not well understood how supramolecular structure and dynamics depend on molecular properties. Steric hindrance induced by molecular entities surrounding the hydroxy group is believed to hamper chain-formation and therefore to influence the Debye process. This was examined by Johari et al.\,\cite{Johari:1972b} in a study of isomeric phenyl-propanols, in which they varied the position of phenyl- and hydroxy group and analyzed the influence on $g_\text{k}$ and spectral shape. In a more recent study it was stated that for 1-phenyl-1-propanol the steric hindrance fully suppresses the structure formation through H-bonding and therefore no Debye contribution is present in the spectra\,\cite{Johari:2001}.

Especially in the case of phenyl-alcohols, but also in many other monoalcohols, it is not easily possible to separate $\alpha$-process and Debye process in the dielectric spectra. This is because contributions of cross-correlations (Debye-like process) and self-correlations overlap and therefore only in the case of strong dynamic separation of both parts a distinction is usually straight forward.  However, it was recently shown for several monoalcohols that photon correlation spectroscopy (PCS) can serve as a tool to identify the self-correlation contributions of the dynamics, because cross correlations were found to either be negligible\,\cite{Gabriel:2017a} or to contribute only very weakly\,\cite{Gabriel:2018a} to the PCS spectra in monoalcohols.

To obtain a deeper understanding of the dynamics in H-bonded systems we combine BDS and PCS to study a series of phenyl-propanols. We apply PCS measurements to identify the $\alpha$-process, which helps us to separate it from the dynamics of supramolecular structures in the BDS spectra. By combining these results with the analysis of the Kirkwood/Fröhlich correlation factor it is possible to obtain more detailed information on the nature of the occurring supramolecular structures and their dynamics. In order to relate these findings to the molecular architecture of the monoalcohols a particular series of phenyl alcohols was chosen, where the position of the phenyl group is varied while the position of the hydroxy group is kept fixed at the end of the carbon backbone.
This also allows us to answer the question, whether increased steric hindrance leads to a small Debye like contribution in the PCS spectra similar to the observations made in secondary alcohols in Ref.\,\cite{Gabriel:2018a}. A graphical overview of the monoalcohols investigated in the present work is given in Fig.~\ref{fig:uebersich7}.


\section{Experimental Background}
Samples of 1-phenyl-1-propanol (1P1P) (Acros Organics, 99\%), 2-phenyl-1-propanol (2P1P) (Aldrich, 97\%) and 3-phenyl-1-propanol (3P1P) (Alfa Aesar, 99\%) were purchased, filtered into a PCS sample-cell by using a $200\,\text{nm}$ hydrophilic syringe filter. The samples for dielectric spectroscopy were used without further purification.

The PCS and BDS experiments were performed with particular care regarding the temperature calibration in the different dielectric and light scattering setups, achieving an overall accuracy of $\pm 0.5\,$K. BDS was performed using a Novocontrol Alpha-N High Resolution Dielectric Analyzer in combination with a time domain dielectric setup, the details of which were reported earlier\,\cite{Rivera:2004a}. After Fourier transforming the time domain data, combined data sets of the complex dielectric susceptibility $\varepsilon^*(\omega)$ were obtained covering the frequency range of $10^{-6} - 10^6\,$Hz.

The PCS experiments were performed in vertical-horizontal (VH) depolarized geometry under a scattering angle of $90^\circ$ in a setup already described earlier in detail\,\cite{Blochowicz:2013a, Gabriel:2015b,Pabst:2017a}. The measured intensity autocorrelation function $g_2(t)=\langle I_{\text{s}}(t)I_{\text{s}}(0)\rangle /\langle I_{\text{s}}\rangle ^2$ can be converted into the electric field autocorrelation function $g_1(t)=\langle E^*_{\text{s}}(0)E_{\text{s}}(t)\rangle/\langle \vert E_{\text{s}}\vert\rangle^2$ by using the Siegert relation for partially heterodyne scattering as described in more detail by Pabst et al.\,\cite{Pabst:2017a}. The amplitude of the fast dynamics $A_\text{fast}$, which is needed in this context and which describes the amplitude of molecular dynamics not captured by the employed correlators, was estimated as $A_\text{fast} = 0.05$.

The correlation function $g_1(t)$ resulting from a light scattering experiment in VH-geometry contains information on molecular reorientation in the sample and thus in principle is comparable to results from BDS. More precisely, however, there are differences, e.g., because in both methods the reorientation of different molecular properties is monitored. In case of BDS the permanent molecular dipole moment is a vectorial quantity, while the anisotropic polarizability in case of PCS is a tensor.
This leads to a different rank $\ell$ of the Legendre polynomial in the respective correlation functions\,\cite{Kremer:2002a,Berne:1976a}:
\begin{equation}
\Phi_\ell(t)=\langle P_\ell(\cos\theta(t))\rangle,
\label{eq:legendre}
\end{equation}
with $\ell=1$ for BDS and $\ell=2$ for PCS. In both cases $\theta(t)$ is the angle between the positions of the respective molecular axis at times 0 and $t$. Assuming that the optical anisotropy and the permanent dipole moment are mainly located at the same molecular entity and the relaxation processes under consideration are isotropic to good approximation, relations between $\Phi_1$ and $\Phi_2$ can be established depending on the motional geometry of the underlying process. For example, in case of a random jump process the correlation function becomes independent of $\ell$\,\cite{Berne:1976a}.

\begin{figure}[t]
\centering
\includegraphics[width=8cm]{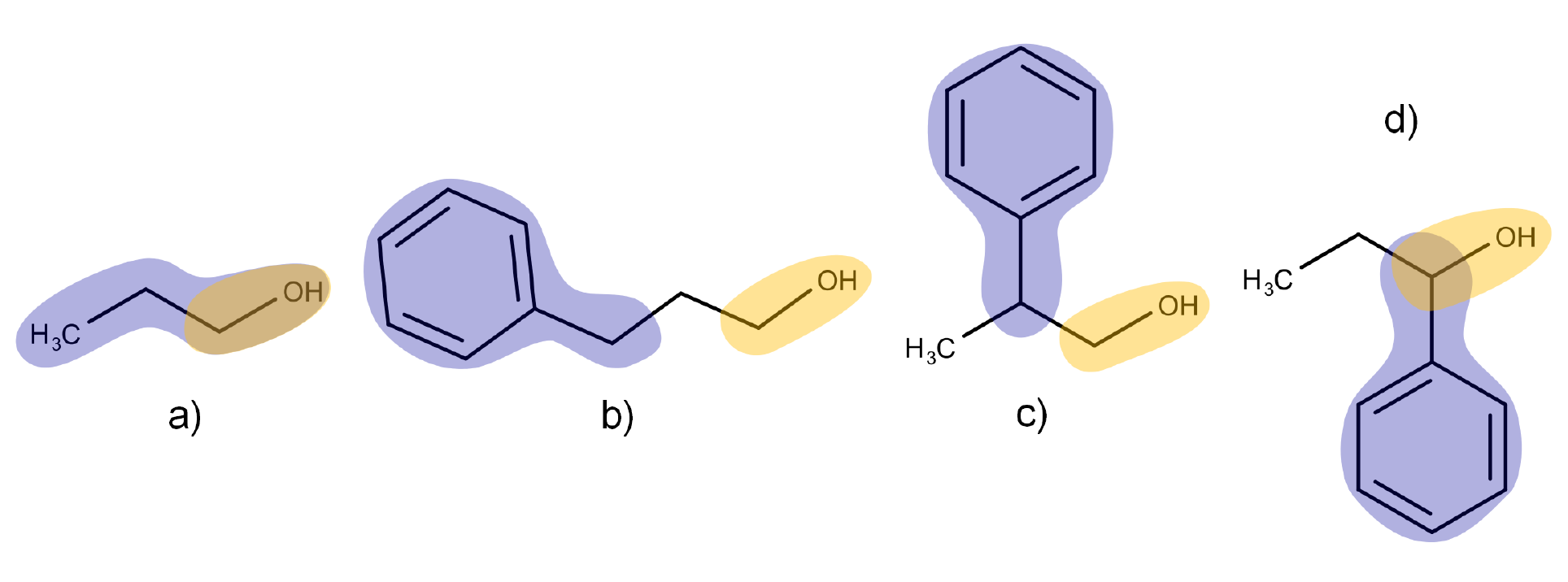}
\caption{Overview of the investigated monoalcohols: (a) 1-propanol, (b) 3-phenyl-1-propanol, (c) 2-phenyl-1-propanol and (d) 1-phenyl-1-propanol. The parts of the molecules containing the permanent dipole moment, measured by BDS, are highlighted in orange and parts carrying the main part of the optical anisotropy and therefore can be examined by PCS are highlighted in blue.} 
\label{fig:uebersich7}
\end{figure}

In Fig.~\ref{fig:uebersich7} an overview of the investigated monoalcohols is given. The parts of the molecules containing the permanent dipole moment measured by BDS are highlighted in orange and the parts that dominate the optical anisotropy monitored in PCS are highlighted in blue. Of course the assignments shown are only schematic. However, calculations of the optical anisotropy parameter yield for 1P $\beta = 1.58$\AA$^3$ and, for 1P1P, 2P1P and 3P1P $\beta = 6.25, 7.42$ and $8.28\,$\AA$^3$ respectively. All values were determined using the program package ORCA\,\cite{ORCA,ORCA2} (DFT functional: B3LYP; basis set: aug-cc-pVTZ). As the scattering intensity $I_{VH}\propto \beta^2$\,\cite{Berne:1976a} these results imply that indeed the phenyl ring dominates the anisotropy in the phenyl alcohols. This is in accordance with experimental observations, which show that the scattering intensity in the phenyl alcohols increases by more than one order of magnitude as compared to 1P. Thus, for 1P the position of both molecular dipole moment and the main part of the optical anisotropy are rather close to good approximation, while in the case of phenyl alcohols the phenyl ring dominates the correlation decay seen in PCS, with the permanent molecular dipole moment, probed by BDS, being located at the hydroxy group in all cases. As a consequence one has to keep in mind that differences may occur between both methods, e.g., if the relaxation geometry shows anisotropic behavior.

Since BDS data are usually treated in frequency domain it is useful for a direct comparison, to obtain PCS data in the same representation. This can be achieved by calculating the imaginary part of the generalized complex susceptibility $\chi''(\omega)$ from $g_1$ through Fourier transformation\,\cite{Kremer:2002a}:
\begin{equation}
\chi''(\omega)=-\omega\int^\infty_0\, g_1(t)\cos(\omega t)\,\mathrm{d}t
\end{equation}
By using the Filon algorithm\,\cite{Filon:1929a} the transformation can be carried out for discrete data sets with logarithmic spacing, resulting in a spectral representation of the light scattering data $\chi''(\omega)$, which is comparable to $\varepsilon''(\omega)$ from BDS. It should be noted that PCS is not able to determine absolute relaxation strengths in a straight forward manner, therefore $\chi''(\omega)$ is usually renormalized when comparing with dielectric data.

Additionally, calorimetric measurements of 1P1P, 2P1P and 3P1P were performed, to determine $T_\text{g}$ and a calorimetric time constant $\tau_\text{cal}$, by using a Perkin-Elmer DSC8000. The measurement program consisted of cycles of cooling from room temperature to $30\,\text{K}$ below the respective $T_{\text{g}}$ with cooling rates of $1$, $2$, $5$, $10$, $20$ and $50\,\text{K}/\text{min}$ and subsequent heating back to room temperature with a fixed heating rate of $20\,\text{K}/\text{min}$. Between every heating and cooling ramp the sample was kept in an isothermal state for $1\,\text{min}$. The heat flow to and from the sample was recorded  during the whole measurement process and analyzed to determine a calorimetric relaxation time as described further below.

\section{Results and Data Analysis}
\begin{figure}[h!]
\centering
\includegraphics[width=8cm]{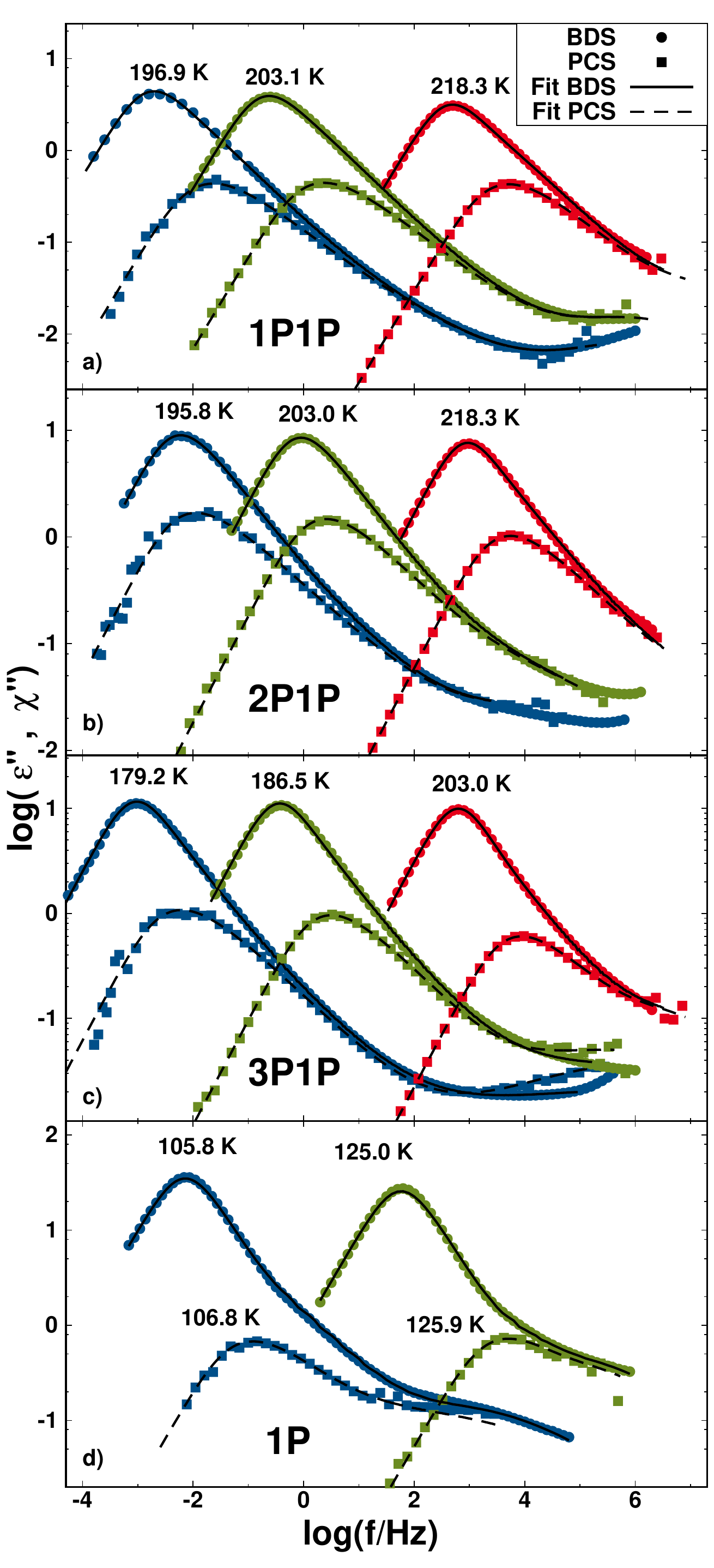}
\caption{Generalized susceptibilities measured in BDS (\ding{108}) and PCS (\ding{110}) of (a) 1-phenyl-1-propanol, (b) 2-phenyl-1-propanol, (c) 3-phenyl-1-propanol and (d) 1-propanol at selected temperatures. Interpolations of BDS (solid lines) and PCS (dashed line) data were performed as described in the text. The normalized PCS spectra are shifted by the factor $\Delta\varepsilon_{\alpha\beta}$.} 
\label{fig:spectra}
\end{figure}

In Fig.~\ref{fig:spectra} we present an overview of PCS and BDS data of 1P1P, 2P1P, 3P1P and 1P at selected temperatures. For better comparison the presented temperatures for the different monoalcohols were chosen such that the relaxation time of the respective main process in PCS approximately matches. Note that the amplitudes of the PCS spectra were shifted according to the data analysis, as discussed further below. The data of 1P and 1P1P were already  presented and discussed previously\,\cite{Gabriel:2017a,Gabriel:2018c}. All investigated monoalcohols were investigated with BDS by other groups before\,\cite{Hansen:1997,Johari:1972b,Johari:2001,Kalinovskaya:2001} and all the previous data are in accordance with our measurements in terms of spectral shape and time constants of the main relaxation peak within experimental accuracy. 

The generalized susceptibilities in Fig.~\ref{fig:spectra} show that the main relaxation processes are faster and more broadened in PCS compared to those in BDS. This is the same behaviour as we have previously shown for other monoalcohols\,\cite{Gabriel:2018,Gabriel:2017a,Gabriel:2018a,Gabriel:2018c}.

To quantitatively analyze and compare the data from PCS and BDS we first interpolate the spectral shape of the PCS measurements.  Assuming the $\alpha$-relaxation to be the same in shape and timescale in both methods, as was already found for several monoalcohols, the corresponding BDS spectrum can then be described by the weighted sum of the interpolated PCS spectrum at the respective temperature, representing the self-correlational contributions, and a single Debye-like relaxation process, representing the cross-correlational contributions. By follwing this procedure spectra of several monoalcohols, including 1P, were successfully analyzed before\,\cite{Gabriel:2018,Gabriel:2017a,Gabriel:2018a,Gabriel:2018c}.

The PCS spectra are interpolated by a superposition of two relaxation processes, which we identify as $\alpha$- and $\beta$-relaxation. Both processes are modeled by using a characteristic distribution of relaxation times, which results in the respective relaxation functions $\Phi_\alpha(t)$ and $\Phi_\beta(t)$. The distributions of relaxation times were chosen following Blochowicz et al.\,\cite{Blochowicz:2003a} as the generalized gamma distribution including a high frequency wing to describe the $\alpha$-process, and a distribution generating an asymmetrically broadened relaxation function to model the $\beta$-process. The combined relaxation function $\Phi_\text{PCS}(t)$ was calculated by using the William-Watts approach
\begin{equation}
  \Phi_\text{PCS}(t) = \Phi_\alpha(t)\cdot\left(\left(1-k_\beta\right) + k_\beta\Phi_\beta(t)\right)
  \label{eq:WW}
\end{equation}
with $k_\beta$ being the relative strength of the $\beta$-process\,\cite{Williams:1971a}. Finally, $\Phi_\text{PCS}(t)$ could be transformed into the frequency domain and used to interpolate the measured PCS spectra by calculating the Fourier transform. Fig.~\ref{fig:spectra} shows interpolations by equation\,(\ref{eq:WW}) of the given PCS data as dashed lines.

To allow for a slight broadening of the additional cross-correlational contribution in the BDS spectra, the latter is desribed by a Kohlrausch-William-Watts (KWW) stretched exponential function with the stretching parameter $0.85 < \beta_\text{KWW,D}\leq 1$. Such a slight broadening of the cross-correlational contribution is expected in cases of weak dynamic separation from the $\alpha$-process, as will be discussed later in more detail. Thus, $\varepsilon''(\omega)$ data from BDS can be interpolated by the Fourier transformation of the following relaxation function:
\begin{equation}
  \Phi_\text{BDS}(t)=\Delta\varepsilon_\text{D}\cdot e^{-\left(\frac{t}{\tau_\text{D}}\right)^{\beta_\text{KWW,D}}} + \Delta\varepsilon_{\alpha\beta}\cdot\Phi_\text{PCS}(t)
  \label{eq:WW2}
\end{equation}
The relaxation strenghts of the cross-correlational and the self-correlational contribution are given by $\Delta\varepsilon_\text{D}$ and $\Delta\varepsilon_{\alpha\beta}$, respectively.  Any differences, which occur between BDS- and PCS spectra regarding shape and strength of secondary relaxation processes were dealt with by constraining the frequency range in which the interpolation with equation (\ref{eq:WW2}) is calculated. The resulting interpolation functions of BDS data are displayed in Fig.~\ref{fig:spectra} represented by the solid lines. The normalized PCS spectra in Fig.~\ref{fig:spectra} are sscaled by the factor $\Delta\varepsilon_{\alpha\beta}$ to display the contribution of $\alpha$- and $\beta$-process in the BDS spectra.

From the BDS data we calculated the Kirkwood/Fröhlich correlation factor\,\cite{Kirkwood:1939a,Froehlich:1958a}:
\begin{equation}
g_\text{k}=\frac{9k_\text{B}\varepsilon_0 MT}{\rho N_\text{A} \mu^2}\frac{(\varepsilon_\text{s}-\varepsilon_\infty)(2\varepsilon_\text{s}+\varepsilon_\infty)}{\varepsilon_\text{s}(\varepsilon_\infty+2)^2},
\label{equ:kirkwood}
\end{equation}
where $k_\text{B}$ is Boltzmann's factor, $\varepsilon_0$ is the permittivity of the vacuum, $M$ the molar mass, $\rho$ the density, $N_\text{A}$ Avogadro's number and $\mu$ is the molecular dipole moment. The static permittivity $\varepsilon_\text{s}$ was directly extracted from the dielectric data, whereas  $\varepsilon_\infty=1.1\,n^2$ was chosen following the literature\,\cite{Johari:1972b}. For that purpose the refractive index $n$ was measured at $532\,$nm in the temperature range of $283-333\,$K by the use of a refractometer and its temperature dependence was extrapolated in a linear fashion to lower temperatures. From the measured refractive index $n(T)$ the density $\rho(T)$ was calculated by using the Lorentz-Lorenz-equation using a molecular polarizability of $16.91\,\text{\r{A}}^3$, which was determined by DFT-calculations. The calculated densities and their temperature dependence agree with literature data within experimental uncertainty. Finally a dipole moment of $1.68\,$D\,\cite{Johari:1972b} and a molecular mass of $136.19\,$g/mol ($60.10\,$g/mol for 1P) was used for the investigated monoalcohols to calculate $g_\text{k}$. The results are displayed in Fig.~\ref{fig:kirkwood}. The error bars were determined by considering the uncertainties of $\varepsilon_\text{s}$, $T$ and $\rho$ alongside a methodic uncertainty in the determination of $\varepsilon_\infty$ (the prefactor of $n^2$ was assumed to be $1.1\pm0.05$).

For the analysis of the calorimetric data the onset glass transition temperatures were determined for 1P1P ($T_\text{g}=197.3\,\text{K}$), 2P1P ($T_\text{g}=197.1\,\text{K}$) and 3P1P ($T_\text{g}=179.2\,\text{K}$).  
Additionally, a calorimetric time constant was calculated in each case by analyzing the fictive temperature $T_\text{f}$ of the glass transition as a function of the applied cooling rate $q$. The fictive temperature $T_\text{f}$ is calculated following the enthalpy matching procedure presented by Moynihan et al.\,\cite{Moynihan:1976}. By interpolating the results with the relation $\ln{q}=A-\Delta H_\text{eff}/RT_\text{f}$, where $A$ is a constant and $R$ is the gas constant, the calorimetric glass transition activation energy $\Delta H_\text{eff}$ can be determined\,\cite{Wu:2017}. The calculated $\Delta H_\text{eff}$ is related to a characteristic calorimetric relaxation time at $T_g$ by\,\cite{Hodge:1994}:
\begin{equation}
\tau_\text{cal}=\frac{RT_\text{g}^2}{q\Delta H_\text{eff}},    
\end{equation}
with $T_\text{g}$ being the onset glass transition temperature and $q$ in our case being a rate of $20\,$K/min.

\section{Discussion}
\begin{figure}[t]
\centering
\includegraphics[width=8cm]{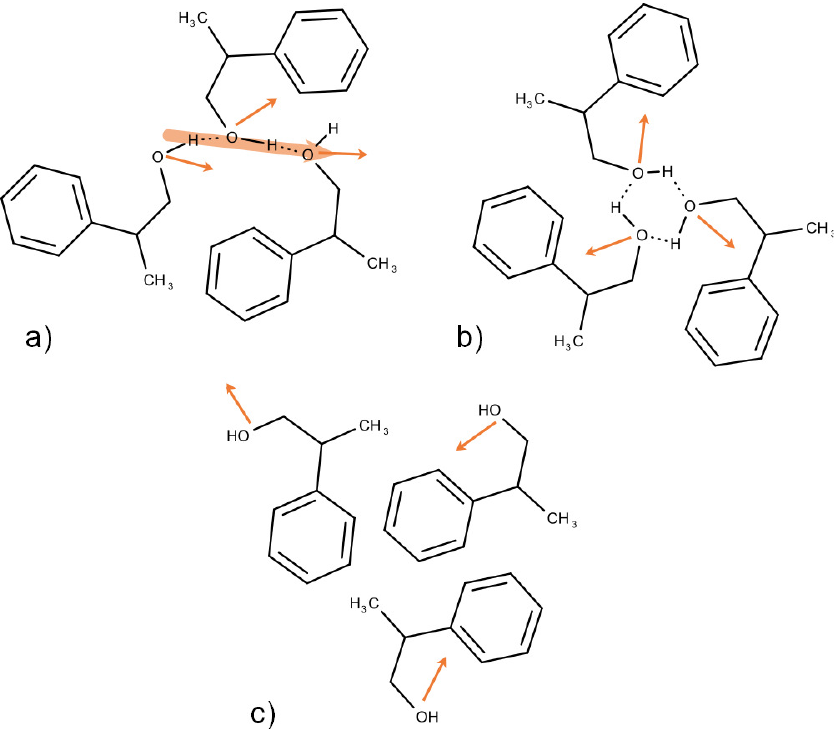}
\caption{Overview of possible supramolecular structures in phenyl alcohols using the example of 2P1P: (a) chain formation, (b) ring formation and (c) suppressed structure formation.} 
\label{fig:uebersich10}
\end{figure}

A basic way of examining a system regarding supramolecular associations is to analyze the static dielectric permittivity of the sample. This can be done by using the Onsager equation\,\cite{Onsager:1938a} corrected by the Kirkwood/Fröhlich correlation factor\,\cite{Kirkwood:1939a,Froehlich:1958a} (see equation (\ref{equ:kirkwood})). The correlation factor describes enhanced or reduced static permittivity generated by orientational cross-correlations between neighboring dipole moments. Fig.~\ref{fig:uebersich10} gives a schematic overview of possible scenarios regarding supramolecular structures in phenyl alcohols. Chain-like (a) or ring-like (b) structures would lead to positive ($g_\text{k}>1$) or negative ($g_\text{k}<1$) orientational cross-correlations between neighboring dipole moments. On the contrary, (c) displays the case of suppression of H-bonds, leading to no significant cross-correlation between dipole moments and therefore $g_\text{k}\approx 1$.

\begin{figure}[t]
\centering
\includegraphics[width=8cm]{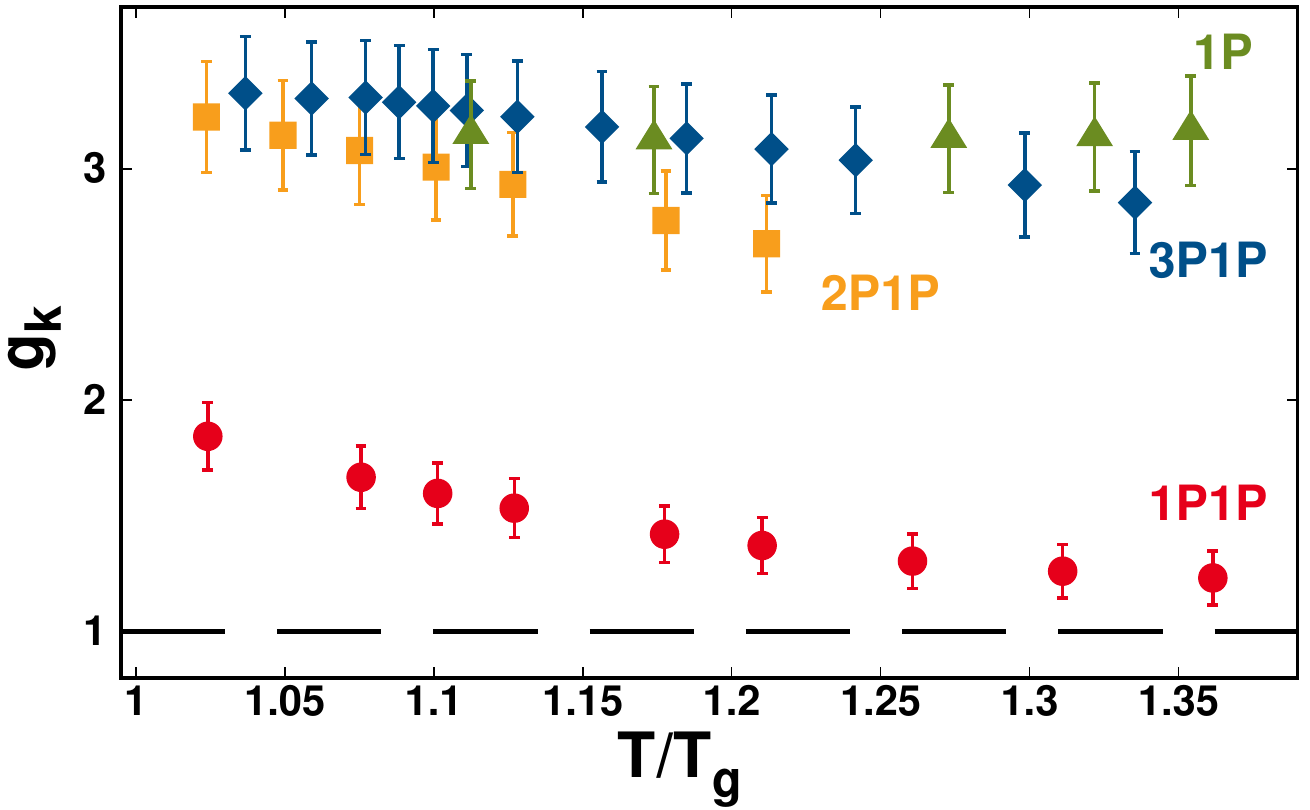}
\caption{Temperature dependent Kirkwood/Fröhlich correlation factor $g_{\text{K}}$ from BDS of 1-phenyl-1-propanol (\ding{108}), 2-phenyl-1-propanol (\ding{110}) and 3-phenyl-1-propanol (\ding{117}) calculated as described in the text. Corresponding data of 1-propanol (\ding{115}) is given for reference. Dashed line is guide for the eye at  $g_\text{k}=1$.} 
\label{fig:kirkwood}
\end{figure}

Fig.~\ref{fig:kirkwood} displays $g_\text{k}(T)$ of 1P, 1P1P, 2P1P and 3P1P around their respective glass transition temperatures. 1P shows a $g_\text{k}$ clearly larger than 1, which already was explained earlier with the occurrence of hydrogen bonded transient chains, leading to an orientational correlation between molecules in the chain\,\cite{Sillren:2014a,Gainaru:2010a}. Both, 3P1P and 2P1P, display a similar behaviour to that of 1P, albeit their $g_\text{k}$ being slightly lower than the one of 1P for most of the temperatures. Following the model of transient chains this indicates an inhibiting effect on H-bonded chain formation induced by the phenyl group, as was already suggested in the literature\,\cite{Johari:2001, Kalinovskaya:2001}. Additionally, $g_\text{k}$ of 2P1P is smaller than the one of 3P1P, allowing the conclusion that a shorter distance between phenyl- and hydroxy group induces a stronger suppression of chain formation. This is also supported by the fact that $g_\text{k}$ of 1P1P, which has the shortest distance between phenyl- and hydroxy group, has the lowest $g_\text{k}$ of all monoalcohols investigated in this work. In this case, however, the situation is slightly different, as 1P1P shows $g_\text{k}\approx 2$ around $T_\text{g}$, but $g_\text{k}$ approaches 1 for higher temperatures. A $g_\text{k}$ in the order of 1 allows for two different possible explanations: Either chain formation is almost fully suppressed by steric hindrance of the phenyl group, or there is a balance between both positive- and negative orientational cross-correlations induced by chain- and ring-like supramolecular structures. Indications for the latter scenario were already reported for different monoalcohols in the literature\,\cite{Dannhauser:1968a,Dannhauser:1968b,Kashtanov:2005,Singh:2012,Young-Gonzales:2016} and can even lead to values of  $g_\text{k}$ being smaller than 1 and also to a crossover from $g_\text{k} > 1$ to $g_\text{k} < 1$ as a function of temperature. We note here, however, that definite conclusion in case of 1P1P becomes difficult due to the calculation of $g_\text{k}$ being highly sensitive to variations in $\varepsilon_\infty$ leading to substantial uncertainties in the values of $g_\text{k}$ as indicated by the error bars in Fig.~\ref{fig:kirkwood}. Hence, some additional analysis considering the entire spectral shape needs to be carried out to allow for further conclusions.

\begin{figure}[t]
\centering
\includegraphics[width=8cm]{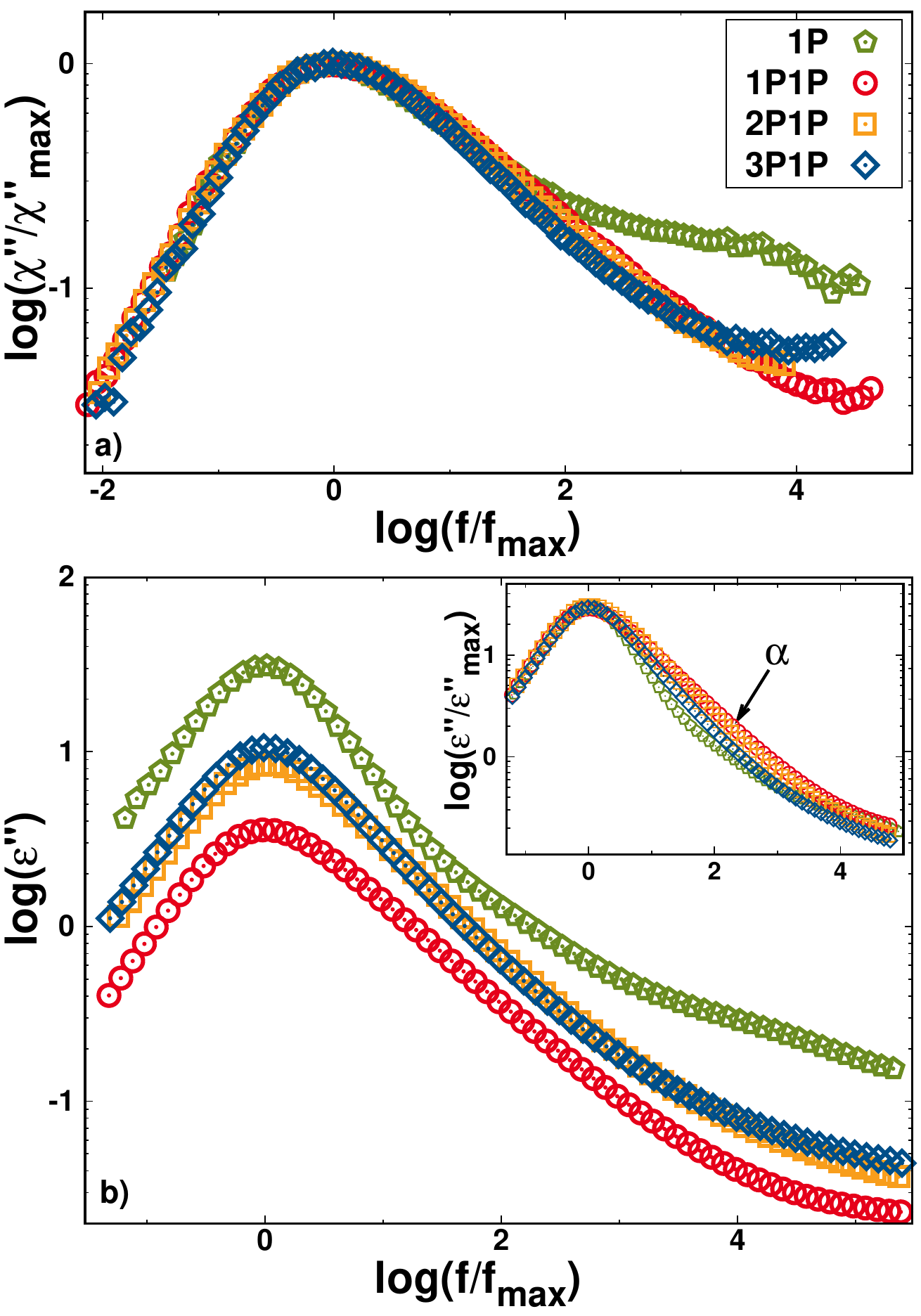}
\caption{Comparison of relaxation spectra of the different investigated monoalcohols measured in PCS (a) and BDS (b) normalized to their respective peak maximum frequency. PCS spectra in (a) and BDS spectra in the inset are additionally normalized to their respective amplitudes.}
\label{fig:shape}
\end{figure}

Fig.~\ref{fig:shape} compares the spectral shape of 1P, 1P1P, 2P1P and 3P1P measured by using (a) PCS and (b) BDS. The data were chosen to approximately match in terms of their respective peak maximum frequency and were normalized to the latter for better comparison. The BDS spectra show strong deviations among each other in terms of spectral shape and amplitude. The differences in spectral shape become more clear by inspecting the inset, where the data were additionally normalized to their peak maximum amplitude: While the spectral shape of all main relaxation processes equals at the low frequency flanks, deviations among the different monoalcohols appear at $f>f_\text{max}$. 

In contrast to this, PCS spectra coincide very well in terms of shape for all three phenyl alcohols and also for 1P if the JG-$\beta$-process at high frequencies is disregarded. Considering the differences of the occurring orientational cross-correlations between the different monoalcohols in BDS (see Fig.~\ref{fig:kirkwood}), we conclude that PCS does not show any dynamic signature of supramolecular structures in the phenyl alcohols. This implies that PCS spectra only display the $\alpha$-process plus additional secondary relaxation processes. Furthermore, the interpolation of the PCS-$\alpha$-process reveals a common spectral shape for 1P and its associated phenyl alcohol isomers. We note that prior studies of the monoalcohols 5-methyl-2-hexanol and 2-ethyl-1-butanol also revealed a very similar $\alpha$-process shape\,\cite{Gabriel:2018a}, which indicates a common $\alpha$-process shape in a broader range of monoalcohols.

For the present analysis this observation is particularly important,  as it indicates that the shape of the $\alpha$-process is identical for all the investigated substances and that the apparent changes in lineshape observed in the dielectric spectra will therefore be due to different contribution of an additional  slow, mostly Debye-like, relaxation process, which superimposes the otherwise common $\alpha$-relaxation. Of course it is non-trivial that processes from $\ell=1$ and $\ell=2$ correlation functions are indeed identical, however this has been shown to be the case for random large-angle jump-like process\,\cite{Gabriel:2018} and even for motional models containing small-angle jumps\,\cite{Diezemann:1998}. Experimentally this observation was already pointed out for several other monoalcohols \cite{Gabriel:2017a,Gabriel:2018a} and is most significant in case of 1-propanol itself, as here $\alpha$- and Debye-like contribution are most easily disentangled in the BDS spectra. We note that while the high frequency flank of the $\alpha$-process seems to coincide between dielectric and light scattering spectra for all the alcohols investigated here, only in case of 3P1P some deviations in the region of the $\beta$-relaxation can be anticipated (cf.\ Fig.~\ref{fig:spectra}c), which is not further discussed here, as the $\beta$-relaxation is hardly resolved in most of the spectra.  

The finding of a common $\alpha$-process shape supports the utilization of the fitting procedure described above, which identifies BDS spectra as a superposition of a common $\alpha$-process and a slow Debye-like contribution for all investigated monoalcohols including 1P1P (see Fig.~\ref{fig:spectra}). This implies that supramolecular structures, relaxing on a slower time scale than the $\alpha$-process, occur in all investigated monoalcohols. Accordingly, in the cases of 3P1P and 2P1P Kirkwood/Fröhlich correlation factor analysis suggests positive orientational cross-correlations and therefore chain-like structures to occur. As a Debye-like contribution can also be identified in case of 1P1P, the observed Kirkwood/Fröhlich correlation factor (see Fig.~\ref{fig:kirkwood}) could be explained by a coexistence of ring- and chain-like structures in this system. Accordingly, the observed convergence to $g_\text{k}\approx1$ approaching higher temperatures can be interpreted as a crossover from the existence of predominantly chain-like to an equilibrium of chain-like and ring-like structures, rather than a suppression of supramolecular structures. We note here, that especially in 1P1P only the combination of light scattering and dielectric measurements allows to disentangle the $\alpha$-process from cross correlation contributions in the dielectric spectra due to the small separation of $\alpha$-process and Debye contribution. Therefore, it is not surprising that previous studies come to a different conclusion regarding the existence of supramulecular structures in particular in 1P1P\,\cite{Johari:2001}. We also note that in principle, one should also consider the possibility of $\pi$-$\pi$-interactions in the formation of supramolecular structures for molecules containing phenyl rings\,\cite{Sato:2005}. However, if the latter played a significant role in our systems some effect in  PCS measurements would be anticipated, which is not observed.

\begin{figure}[t]
\centering
\includegraphics[width=8cm]{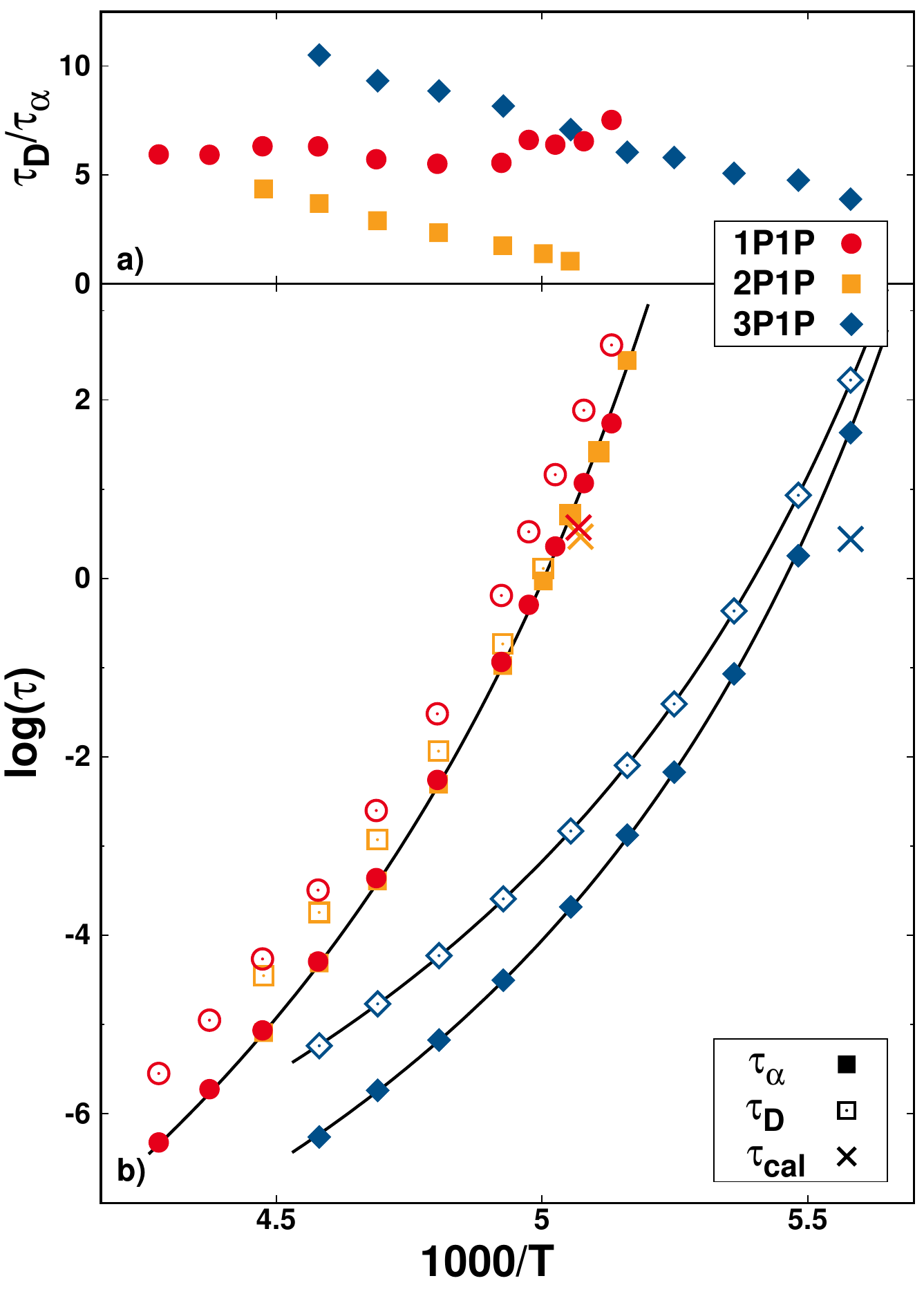}
\caption{Lower plot (b) shows BDS time constants of $\alpha$- and Debye process and a single calorimetric time constant of 1P1P, 2P1P and 3P1P in dependence of the inverse temperature. Upper plot (a) displays the dynamic separation $\tau_\text{D}/\tau_\alpha$ of $\alpha$- and Debye process.} 
\label{fig:timeconstant}
\end{figure}

Additional insight can be gained by considering the temperature dependence of the different relaxation processes. Fig.~\ref{fig:timeconstant} (b) displays the time constants of the $\alpha$-process $\tau_\alpha$ and the Debye-process $\tau_\text{D}$ of the investigated phenyl-alcohols. As expected, both processes follow a Vogel-Fulcher-Tamann (VFT) temperature dependence, however only selected interpolations are displayed for reasons of clarity. Additionally, a single calorimetric time constant $\tau_\text{cal}$ is plotted for every phenyl-alcohol. For both 1P1P and 2P1P $\tau_\text{cal}$ matches with the time constants of the respective $\alpha$-process in good agreement, while being different from $\tau_\text{D}$, especially in the case of 1P1P.  Thus, there seems to be no or only weak signature of supramolecular structures in the calorimetric data, which corresponds to the behavior reported in previous studies\,\cite{Huth:2007}. In the case of 3P1P, $\tau_\text{cal}$ is clearly faster than $\tau_\alpha$, for yet unknown reasons. This decoupling of calorimetric and dynamic $\alpha$-process is unexpected and requires further investigation. To achieve a better comparison of calorimetric and dynamic measurements it would therefore be useful to analyze the dynamic heat capacity $c_\text{p}^*(\omega)$ and verify the occurrence of the mentioned decoupling\,\cite{Huth:2007}.

In addition, Fig.~\ref{fig:timeconstant} (a) displays the temperature dependence of the dynamic separation between $\alpha$-process and Debye-process, calculated by the quotient of $\tau_\text{D}$ and $\tau_\alpha$. In the cases of 2P1P and 3P1P the dynamic separation is decreasing towards lower temperatures. Similar results were found in the studies of Bauer et al.\,\cite{Bauer:2013}, where for several monoalcohols $\tau_\text{D}/\tau_\alpha$ approaches a maximum at $\tau_\alpha\approx10^{-4}\,$s and decreases for larger values of $\tau_\alpha$. In contrast to that $\tau_\text{D}/\tau_\alpha$ stays roughly constant over the investigated temperature range in the case of 1P1P. This might result from the occurrence of ring-like structures, or even from the change of the amount of ring- and chain-like associations over temperature. Further investigation is needed to develope a more detailed picture of how dynamic separation and its temperature dependence behave for different monoalcohols.

One striking feature of the Debye process in monoalcohols is its association with only one single relaxation time. This characteristic feature is usually explained in the framework of the transient chain model when considering the different dynamic timescales in the system \cite{Gainaru:2010,Gainaru:2011}: As the relaxation of the transient chain structure takes place on a timescale ($\tau_\text{D}$) longer than the timescale of the fluctuations of the surrounding environment ($\tau_\alpha$) the transient chain as a larger object effectively averages over the surrounding dynamic heterogeneities, leading to rate exchange faster than the slow relaxation time and thus only to a single timescale $\tau_\text{D}$ \cite{Anderson:1967,Boehmer:2014a}. However, phenyl alcohols, considering the absolute values of $\tau_\text{D}/\tau_\alpha$, usually display only a small dynamic separation between $\alpha$- and Debye process in comparison to many other monoalcohols. Therefore this averaging over the dynamic heterogeneities is sometimes incomplete and accordingly a certain broadening of the slowest relaxation process can be observed.

\begin{figure}[H]
\centering
\includegraphics[width=8cm]{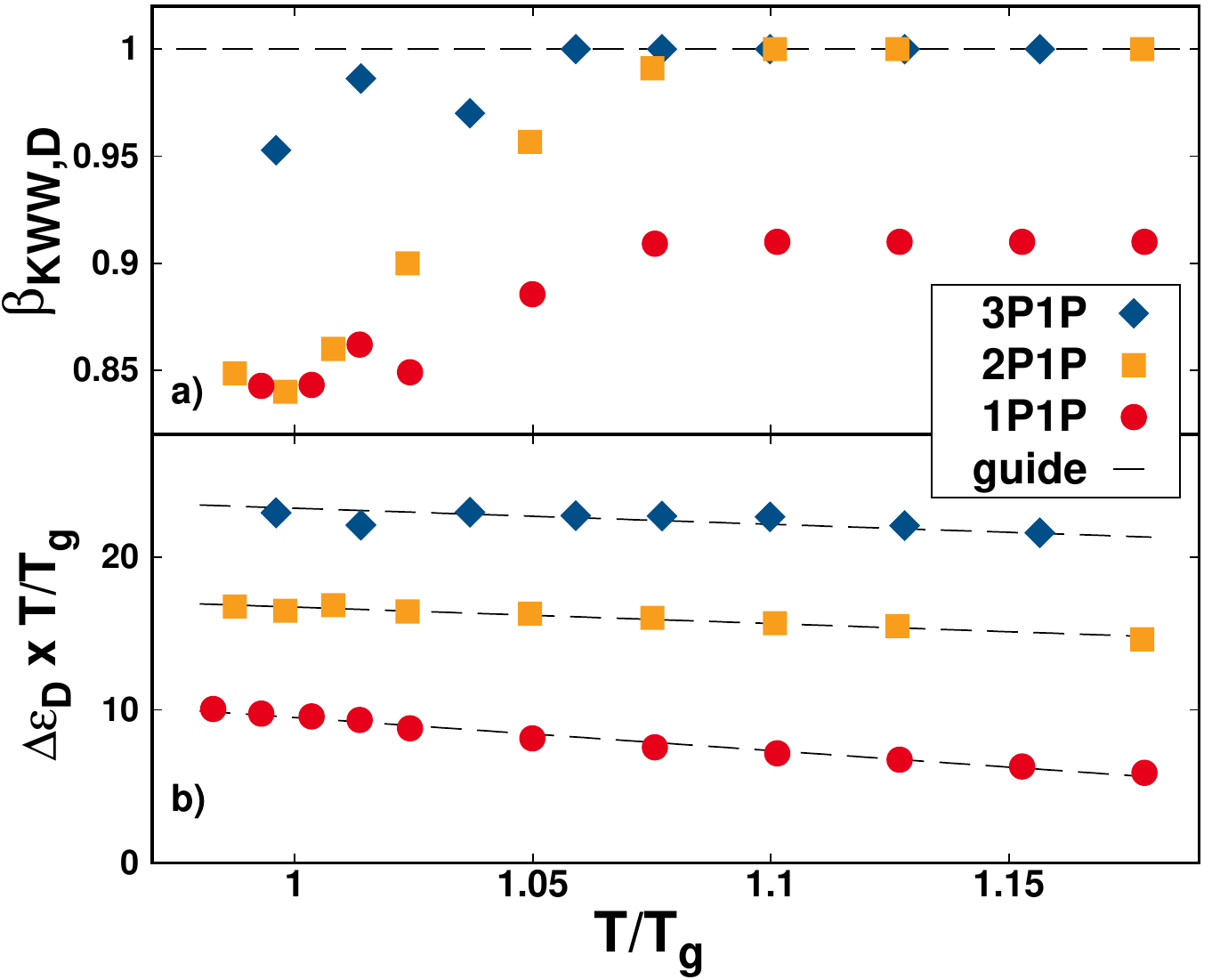}
\caption{Upper plot (a) displays the temperature dependent broadening parameter $\beta_\text{KWW,D}$ of the KWW function describing the Debye process in the spectra of 1P1P, 2P1P and 3P1P. The dashed line is a guide at $\beta=1$, representing a relaxation process with only a single relaxation time. Lower plot (b) shows the temperature dependence of the Debye process relaxation strength $\varepsilon_\text{D}$ with the trivial Curie temperature dependence being extracted.} 
\label{fig:breite}
\end{figure}

To deal with this possibility, the analysis of the slow relaxation process in the dielectric spectra was performed using a KWW stretched exponential function with a stretching parameter $\beta_\text{KWW,D}$ (see Equation (\ref{eq:WW})). Fig.~\ref{fig:breite}\,(a) displays the temperature dependence of $\beta_\text{KWW}$ for the investigated phenyl alcohols. In the case of 2P1P and 3P1P $\beta_\text{KWW}=1$ for high temperatures, which is also the temperature range showing the largest dynamic separation of $\alpha$- and Debye process. At low temperatures the dynamic separation gets smaller and our analysis reveals a slight broadening for 3P1P ($\beta_\text{KWW}\approx 0.95$) and an increased broadening for 2P1P ($\beta_\text{KWW}\approx0.85$). This also coincides with the absolute values of $\tau_\text{D}/\tau_\alpha$, since 3P1P generally shows a larger dynamic separation than 2P1P, which might indicate a decreasing average number of molecules forming transient chains as the phenyl ring is closer to the OH group an steric hindrance towards chain formation increases. For 1P1P the situation is apparently more complicated since the slowest process shows a certain broadening $\beta_\text{KWW}=0.9$ even at high temperatures. However, the strength of the Debye-like relaxation systematically decreases from 3P1P to 1P1P (see Fig.~\ref{fig:breite}\,(b)) as one would expect following the reasoning given above. 

\section{Conclusions}
By comparing photon correlation and dielectric spectroscopy in a systematic series of 1-propanol and various phenyl propanols, it was possible to disentangle the contributions of $\alpha$-relaxation and a slow, Debye-like process ascribed to cross correlations due to the formation of supramolecular structures.  The observed $\alpha$-process displays a common shape for 1P and its associated phenyl alcohol isomers and no Debye-like contribution is observed in PCS. This finding is in agreement with similar observations in other primary alcohols, where also no Debye contribution is observed in PCS. By direct comparison with PCS data at the same temperature, a slow Debye-like process was identified in the BDS sepctra of all investigated phenyl alcohols. The analysis of the  Kirkwood/Fröhlich correlation factor reveals the cross correlation contributions due to supramolecular structures becoming smaller, the closer the phenyl ring is positioned to the hydroxy group, in accordance with the analysis of the dielectric strength of the slow Debye-like peak. In case of 1P1P the analysis furthermore indicates that ring and chain like supramolecular structures coexist and lead to a finite strength of the Debye-like contribution while the correlation factor is close to unity.

Calorimetric measurements support these findings and reveal a $\tau(T_g)$ which coincides as expected with the $\alpha$-relaxation times determined from PCS and BDS. For yet unknown reasons only 3P1P shows deviations from this behavior. Finally the slow Debye-like dielectric relaxation was analyzed in terms of spectral shape, revealing a very slight broadening in particular at low temperatures in the cases of 2P1P and 3P1P, which correlates with the decrease of the dynamic separation of $\alpha$- and Debye-like process. This is the expected behaviour if the relaxation of the supramolecular structures is hardly slower than the $\alpha$-relaxation. Thus, it turns out the cross correlation contributions play a major role in all of the dielectric spectra of all of the phenyl alcohols  and only a systematic comparison of BDS and PCS data allows to clearly disentangle both contributions, which is impossible with only one of the methods alone. If a similar strategy can be applied also in case of polyalcohols or even in case of non hydrogen bonding liquids is subject of further investigations.

\section{Acknowledgement}
Financial support by the Deutsche Forschungsgemeinschaft under Grants
No. BL 923/1 and BL 1192/3 and within FOR 1583  is gratefully acknowledged.

\end{document}